\documentstyle[12pt,axodraw]{article}
\begin{document}
\baselineskip=6.0mm
\begin{titlepage}
\begin{flushright}
  KOBE-TH-99-08 \\
\end{flushright}
\vspace{2.3cm}
\centerline{{\large{\bf Effective Action Approach to Heavy Particle
Contributions}}}
\centerline{{\large{\bf and}}}
\centerline{{\large{\bf Wilsonian Renormalization}}}
\par
\par
\par\bigskip
\par\bigskip
\par\bigskip
\par\bigskip
\par\bigskip
\renewcommand{\thefootnote}{\fnsymbol{footnote}}

\centerline{{\bf C.S.
Lim}$^{(a)}$\footnote[1]{e-mail:lim@phys.sci.kobe-u.ac.jp}
 and {\bf Bungo
Taga}$^{(b)}$\footnote[2]{e-mail:taga@octopus.phys.sci.kobe-u.ac.jp}}
\par
\par\bigskip
\par\bigskip
\centerline{$^{(a)}$ Department of Physics, Kobe University, Nada, Kobe
657-0013, Japan}
\centerline{$^{(b)}$ Graduate School of Science and Technology, Kobe
University, Kobe 657-0013, Japan}
\par
\par\bigskip
\par\bigskip
\par\bigskip
\par\bigskip
\par\bigskip
\par\bigskip
\par\bigskip
\par\bigskip
\par\bigskip
\centerline{{\bf Abstract}}\par

We work in theories with both light and heavy particles. A method to
obtain an effective low energy action with respect to the light particle is
presented. Thanks to Wilsonian renormalization, we obtain effective
actions with finite number of local operators describing possible
non-decoupling effects of heavy particles. The validity of the method is
first checked by explicit computation of a specific observable, the $\rho$ -
parameter, in the framework of effective theory. Then we discuss a procedure
to obtain the  full effective action with respect to light particles in a
typical
example of the system of (t, b) doublet, gauge bosons and Higgs doublet,
regarding only t quark as a heavy particle.

\par\bigskip
\par\bigskip
\par\bigskip
\par\bigskip
\par\bigskip
\par\bigskip
\par\bigskip

\end{titlepage}
\newpage

\leftline{\bf 1. Introduction}
\vspace{0.2 cm}

The Standard Model of elementary particle interactions is very successful,
except for failing to explain recently claimed neutrino oscillation
\cite{Kamiokande}.
The model is expected to be an effective low energy theory of
some more fundamental theory ("new physics") with a larger symmetry, such as
Minimal Supersymmetric Standard Model(MSSM), Grand Unified Theory (GUT),
super-string, ... The fact that the Standard Model is doing well in lower
energies suggests that new heavy particles, characteristic to each "new
physics",
(superpartners, lepto-quark,...) do not practically affect the low energy
theory or their effects are rather limited. At least in the type of new
physics models where new heavy particles
are expected to be
decoupled from low energy world as in the case of superpartners, the
effective low energy action should be well described by what we obtain just
eliminating
such heavy particles from the original action, though  bare couplings and
field normalizations get some modification: decoupling theorem
\cite{Decoupling}.
Even in the type of new physics models  where new heavy particles do affect
low energy world (non-decoupling), as in the case of a model with heavy
quark of fourth  generation if it ever exists, the non-decoupling radiative
corrections due to heavy particles are expected to be described by
limited number of independent effective operators. For
instance we know that only S,T and U parameters are enough to parametrize
the
non-decoupling effects on gauge boson self energies.

Our purpose in this work is to propose a method to construct the effective
low
energy action with respect to light particles, describing radiative
corrections  due to heavy particles,  in theories with both light and heavy
particles.  Though usually non-decoupling effects of heavy particles are
independently analyzed
for each observable, it will be nice if we can get the full effective low
energy action in a systematic way, since such obtained effective action
should contain
all informations of the possible non-decoupling effects. By the end of this
article, we would like to show by taking a prototype model with only (t,b)
doublet,
gauge bosons and Higgs doublet that this project is realized: we construct
full  effective action concerning gauge bosons, Higgs doublet and light b
quark, which results from the path-integrals of fermionic fields.

After explaining the outline of our method invoking Wilsonian
renormalization,
we check the validity of our method by focusing on the operators in the
effective action which are
relevant for a specific observable, i.e. the $\rho$- parameter. According to
our method we obtain the necessary operators in two
prototype models and calculate the  $\rho$- parameter by use of these
operators, to confirm that they reproduce the known results. Finally we
demonstrate
that to obtain the full effective action in a compact closed form is
possible,
 provided we focus on the non-decoupling effects ($\Delta \rho \equiv \rho -
1$ is one of them \cite{Veltman}).

Key ingredient of our method is to utilize "Wilsonian renormalization"
\cite{Wilson}. (In the present argument, Wilsonian renormalization just
means to integrate out the higher momentum (short distance) part of the
fields or to reduce the
U.V. cutoff of the theory, and R.G. technique will not be used.) The
usefulness of Wilsonian renormalization may be understood as follows. A
natural guess to obtain the effective theory is to (path-)integral  the
heavy particles out from the theory, as heavy particles never appear in the
external lines of low energy processes.
It, however, turns out that such manipulation results in the appearance of
non-local operators with respect to light particles. To see the situation,
suppose that light and heavy particles, denoted by $l$ and $h$,
respectively, are interacting via
$L_{int} = \frac{\lambda}{2} hll$ (with $\lambda$ being a coupling
constant). The path-integral is equivalent to calculating Fig.1 and will
produce effective lagrangian in momentum space,
$\frac{\lambda^{2}}{8}\frac{1}{m_{h}^{2} - k^2} l^4$
($k$: momentum carried by $h$), leading to a non-local operator in
configuration space,
\begin{equation}
L_{eff}
=  \frac{\lambda^{2}}{8} l^2 \frac{1}{m_{h}^{2} + \Box} l^2
= \frac{\lambda^{2}}{8} (\frac{1}{m_{h}^{2}} l^4 - \frac{1}{m_{h}^{4}} l^2
\Box  l^2 + .....),
\end{equation}
where $m_h$ is the mass of $h$. One may naively expect that taking a limit,
$m_h \rightarrow \infty$, the $L_{eff}$ disappears anyway and the
non-locality is not a problem. Unfortunately it is not the case, as $k^2$
can easily exceeds $m_{h}^{2}$ in a Feynman diagram where light particle $l$
participates in the internal loop (Fig.2). In other words, the above
mentioned limit and the integral of loop
momentum are not commutative \cite{Feruglio} (of course, right answer should
be obtained when we first perform loop integral and take the limit
afterwards). Real problem is that even in a theory where decoupling of heavy
particle is anticipated, the decoupling is not manifest at all, because of
the presence of such non-local operator. In the above example, unless
$\lambda$ is correlated with $m_h$, $h$ particle should be decoupled from
the low energy world according to the
decoupling theorem \cite{Decoupling}, though $L_{eff}$ cannot be simply
neglected even for the limit of large $m_h$.


\begin{center}
\begin{picture}(300,100)(0,0)
\Line(10,85)(30,60)
\Line(10,35)(30,60)
\Line(30,60)(70,60)
\Vertex(30,60){2.5}
\Vertex(70,60){2.5}
\Line(70,60)(90,85)
\Line(70,60)(90,35)
\Text(17,85)[]{${l}$}
\Text(17,35)[]{${l}$}
\Text(83,35)[]{${l}$}
\Text(83,85)[]{${l}$}
\Text(50,53)[]{${h}$}
\Vertex(205,65){2.5}
\Line(180,65)(205,65)
\Line(255,65)(280,65)
\Vertex(255,65){2.5}
\CArc(230,65)(25,0,360)
\Text(190,58)[]{${l}$}
\Text(270,58)[]{${l}$}
\Text(230,98)[]{${l}$}
\Text(230,32)[]{${h}$}
\Text(50,15)[]{\sl Fig.1}
\Text(230,15)[]{\sl Fig.2}
\end{picture} \\
\end{center}


The situation would have changed if loop integral from the U.V. cutoff
$\Lambda$ to some scale $\mu$, which is sufficiently smaller than $m_h$
($\mu \ll m_h$ ), had been performed concerning the light particle
(Wilsonian renormalization), i.e. if the U.V. cutoff of the theory had been
reduced from  $\Lambda$ to
$\mu$. After such Wilsonian renormalization, the limit $m_h \rightarrow
\infty$ can be safely taken and the problem of non-locality will disappear.
Thus the decoupling is expected to become manifest. This is why we wish to
invoke Wilsonian  renormalization.

In the literature there is a long list of works discussing effective action
 formalism to deal with radiative corrections (see for instance
\cite{Weinberg},\cite{Cohen},\cite{Feruglio}). Some of them
\cite{Cohen},\cite{Feruglio} are
clearly related with our present work. Their prescriptions, however, do not
utilize Wilsonian renormalization and are more oriented to the structure of
low energy theory. In our approach the effective theory is more directly
constructed from the original theory.

\vspace*{0.5 cm}
\leftline{\bf 2. Low energy effective action and Wilsonian renormalization}

To illustrate our method to obtain an effective action $S_{eff}$, we
consider a  simple model with a pair of light and heavy scalar particles,
$l$ and $h$, with masses $m_l$ and $m_h$, $m_l \ll m_h$. The partition
function is given by a path-integral,
\begin{equation}
Z
=  \int [{\cal D}h][{\cal D}l] e^{iS[h,l]},
\end{equation}
where $S$ is the original action of the system. Our aim is to derive low
energy ($E \ll m_h$) effective action (lagrangian)  which is valid at the
energy range $0 \le E \le \mu$, where the fixed mass scale $\mu$ is the
physical U.V. cutoff of the effcetive low energy theory and taken to be
sufficiently smaller than $m_h$, i.e. $m_l \ll \mu \ll m_h$. For instance,
$m_h \sim 10^{16} GeV$ and $\mu$
 can be taken to be,  say, $ \sim  10^{15}GeV$, and $m_l \le
10^2 GeV$, if the original theory is GUT type theory. For such purpose, we
divide each field into lower and higher momentum parts.
Namely $h = h_{<}+h_{>}$ and $l = l_{<} + l_{>}$, with $h_{<}$ and $h_{>}$
denoting heavy fields with momentum $|k| < \mu$ and $|k| > \mu$ ($|k| \equiv
\sqrt{|k^{\mu}k_{\mu}|}$, after Wick rotation), etc..

Our method based on the Wilsonian renormalization is to perform
path-integrals   with respect to $h_{<}$, $h_{>}$ and $l_{>}$, so that we
can get an effective action  for $l_{<} $ alone:
\begin{equation}
Z
=  \int [{\cal D}h_{<}][{\cal D}h_{>}][{\cal D}l_{<}][{\cal D}l_{>}] e^{i
S[h_{<}+h_{>},l_{<} + l_{>}]}
=  \int [{\cal D}l_{<}] e^{i S_{eff}[l_{<}]},
\end{equation}
i.e., the effective action is given by
\begin{equation}
e^{i S_{eff}[l_{<}]}
=  \int [{\cal D}h_{<}][{\cal D}h_{>}][{\cal D}l_{>}] e^{i
S[h_{<}+h_{>},l_{<} + l_{>}]}.
\end{equation}
At the right-hand side of the above equation the ordering of the
path-integrals  should be such that the integrals of higher momentum parts,
$h_{>}$, $l_{>}$,
are done before the integral of $h_{<}$,
since if we perform $h_{>}$ and $h_{<}$ integrals first the problem of
non-locality arises again.
As the matter of fact, we will see that the path integral of $h_{<}$ always
gives negligible contribution to $S_{eff}$ and the only Wilsonian
renormalization, i.e. the integrations of $h_{>}$ and $l_{>}$, yields
approximately correct
answers of the effective action.

Actually the effective action $S_{eff}$ can be calculated in two ways.
One way is to directly  perform  path-integral, as
we will do in constructing the full effective action. Another way is to
calculate
Feynman diagrams perturbatively. In the latter case, we may assign a double
line for a field with higher momentum and a single line for a field with
lower
momentum and just calculate diagrams where the internal lines are either
double line of heavy or light field or single line of heavy field and
external lines are all single lines of light fields.

A potentially serious problem of our method is concerning local gauge
invariance. In our approach, local gauge invariance may be explicitly broken
by the presence of
$\mu$ and U.V. cutoff $\Lambda$. Actually the breakdown of gauge invariance
due to $\mu$ is not a real problem, as the final result of some observable
should not depend on the choice of $\mu$. A real problem might be caused by
the presence of $\Lambda$. We, however, note that what we are really
interested in are non-decoupling radiative corrections of heavy particles,
which are all described by gauge invariant irrelevant (with mass dimension
$d > 4$) operators provided we include Higgs field as well as
light particles to form operators (the effects described by marginal or
relevant operators can be absorbed in the renormalization of bare couplings
and
fields). The coefficients of the irrelevant operators are automatically
finite and will not suffer from the problem originated from the
regularization by $\Lambda$. For
example, $\Delta \rho$ is described by a $d = 6$ operator including Higgs,
though superficially it is an observable related with $d = 2$  operators,
i.e. gauge boson mass-squared.

\vspace*{0.5 cm}
\leftline{\bf 3.Effective action and the $\rho$-parameter}

To check the validity of our method and to get some insight into the
procedure
toward the full construction of the effective action, we now derive a part
of the effective action which is relevant for a specific observable, $\Delta
\rho = \rho - 1$ or
T-parameter \cite{Peskin}, which is well-known as a typical example of the
non-decoupling effect of heavy particle. The resultant effective action is
then used to calculate the $\Delta \rho$, to confirm that it recovers the
known results.

We take two typical examples of new physics contribution which give
non-decoupling and decoupling effects on
$\Delta \rho$: (a) the contribution of t quark in a SU(2) doublet (t,b) in
the Standard Model, (b) the contribution of a SU(2) doublet of superpartners
of light quarks, ($\tilde{u}$,$\tilde{d}$), in MSSM.

\noindent (a) Doublet (t,b) in the Standard Model

In this case the roles of $h$ and $l$ particles are played by $h$: t quark,
$l$: b quark and SU(2) gauge bosons $W^{+}_{\mu}, W^{-}_{\mu}, W^3_{\mu}$,
respectively. All our discussions in this work are restricted to the 1-loop
level and
for a while the presence of Higgs doublet is ignored except the role of
providing quark and gauge boson masses. We assume $m_b, M_W \ll \mu \ll
m_t$, though t quark is not actually so heavy. If necessary (t,b) doublet
might be understood as the one of fourth generation, for instance. We adopt
the way of calculating Feynman diagrams in order to derive the effective
action which stems from the radiative corrections of $t_{>}$, $b_{>}$ and
$t_{<}$ fields, the part relevant for $\Delta \rho$. The contributing
diagrams up to 1-loop level are shown in Fig.3. For the intermediate double
lines the loop integral is over the momentum range $|k| \ge \mu$, while for
intermediate single line of t quark the loop integral is over the range $|k|
\le \mu$.
The diagrams in Fig.3 are calculated to provide an effective lagrangian,
\begin{eqnarray}
L_{eff}
=  (C_1 + \frac{1}{2}C_2) W^{+}_{\mu}W^{-\mu} +
\frac{1}{2}(C_1 - \frac{1}{2}C_2) W^{3}_{\mu}W^{3\mu} \nonumber \\
+ \frac{g^2}{2m_t^2} \bar{b}W_{\mu}^{-}\gamma^{\mu}
i(\gamma^{\lambda}\partial_{\lambda}) W^{+}_{\nu}\gamma^{\nu}
\frac{1-\gamma_5}{2}  b .
\end{eqnarray}
The divergent coefficient $C_1$ is common for both charged and neutral gauge
bosons and can be absorbed into redefinition of bare gauge boson masses 
in the original lagrangian. The
coefficient $C_2$
gets finite contributions from the combination of Fig.3a and 3b and from
Fig.3c:
\begin{equation}
C_2 = C_2^{(3a+3b)} + C_2^{(3c)},
\end{equation}
where
\begin{eqnarray}
C_2^{(3a+3b)} &\simeq& \frac{3g^2}{64\pi^2}(m_t^2 - \mu^2), \nonumber \\
C_2^{(3c)}    &\simeq& \frac{-g^2}{64\pi^2} \frac{\mu^6}{m_t^4}.
\end{eqnarray}
In the above expression $m_b$ has been ignored and among the $\mu$ dependent
terms only the leading term of the expansion in $\mu^2/m_t^2 \equiv r \ll 1$
has been kept in each contribution.
The remaining term in the second line of eq.(5) comes from the tree diagram
Fig.3d. As emphasized in the introduction, thanks to the Wilsonian
renormalization, taking only the leading term of the expansion in
$|\Box/m_t^2| \leq
\mu^2/m_t^2 = r$ can be
justified when we evaluate Fig.3d, thus making $L_{eff}$ a set of finite
number of local operators.

\begin{center}
\begin{picture}(385,70)(0,0)
\Text(40,60)[]{\scriptsize ${W^+}$}
\Text(90,60)[]{\scriptsize ${W^-}$}
\Text(63,70)[]{\scriptsize ${t}$}
\Text(67,30)[]{\scriptsize ${b}$}
\Photon(30,50)(50,50){3}{3}
\Photon(80,50)(100,50){3}{3}
\CArc(65,50)(15,0,360)
\CArc(65,50)(13,0,360)
\Text(65,10)[]{(a)}
\Text(135,60)[]{\scriptsize ${W^3}$}
\Text(185,60)[]{\scriptsize ${W^3}$}
\Text(158,70)[]{\scriptsize ${t,b}$}
\Text(162,30)[]{\scriptsize ${t,b}$}
\Photon(145,50)(125,50){3}{3}
\Photon(195,50)(175,50){3}{3}
\CArc(160,50)(15,0,360)
\CArc(160,50)(13,0,360)
\Text(160,10)[]{(b)}
\Text(230,60)[]{\scriptsize ${W^3}$}
\Text(280,60)[]{\scriptsize ${W^3}$}
\Text(255,70)[]{\scriptsize ${t}$}
\Text(255,30)[]{\scriptsize ${t}$}
\Photon(220,50)(240,50){3}{3}
\Photon(270,50)(290,50){3}{3}
\CArc(255,50)(15,0,360)
\Text(255,10)[]{(c)}
\Photon(325,65)(340,50){3}{3}
\Photon(375,65)(360,50){3}{3}
\ArrowLine(340,50)(360,50)
\ArrowLine(325,35)(340,50)
\ArrowLine(360,50)(375,35)
\Text(320,55)[]{\small  $ W^- $}
\Text(375,55)[l]{\small $ W^+ $}
\Text(330,35)[l]{$ b $}
\Text(370,35)[r]{$ b $}
\Text(350,57)[]{$ t $}
\Text(350,10)[]{(d)}
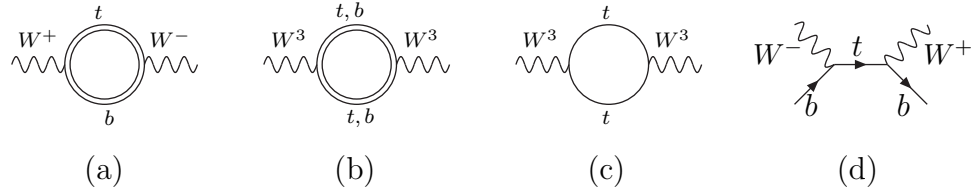
\end{picture}    \\{\sl Fig.3}
\end{center}

\begin{center}
\begin{picture}(300,100)(0,0)
\Photon(170,50)(200,50){3}{4}
\Photon(250,50)(280,50){3}{4}
\CArc(225,50)(25,0,360)
\Text(180,60)[]{$ W^3 $}
\Text(220,80)[]{$ b $}
\Text(230,20)[]{$ b $}
\Text(225,0)[]{(b)}

\Photon(30,25)(120,25){3}{12}
\CArc(75,48)(23,0,360)
\Text(40,35)[]{$ W^+ $}
\Text(110,35)[]{$ W^- $}
\Text(75,80)[]{$ b $}
\Text(75,0)[]{(a)}
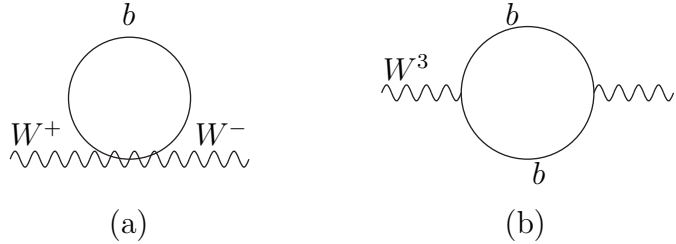
\end{picture}     \\{\sl Fig.4}
\end{center}


 Let us calculate $\Delta \rho$ by use of the derived effective lagrangian.
In addition to the direct contributions from  $C_2^{(3a+3b)}$ and
$C_2^{(3c)}$, there are indirect contributions from  the term in the second
line of eq.(5) and from ordinary neutral current interaction of $b_{<}$
field in the
original lagrangian, through 1-loop diagrams shown in Fig.4a and Fig.4b,
respectively:
\begin{equation}
\Delta \rho = \frac{1}{M_W^2}(C_2^{(3a+3b)} + C_2^{(3c)} + C_2^{(4a)} +
C_2^{(4b)}),
\end{equation}
with
\begin{eqnarray}
C_2^{(4a)} &\simeq& - \frac{3g^2}{64\pi^2} \frac{\mu^4}{m_t^2}, \nonumber \\
C_2^{(4b)} &\simeq& \frac{3g^2}{64\pi^2} \mu^2.
\end{eqnarray}
$C_2^{(4a)}$ and $C_2^{(4b)}$ denote the contributions from Fig.s 4a and 4b,
respectively.
In the summation of  all contributions the leading $\mu$ dependent term,
$\mu^2$ term, cancel out between  $C_2^{(3a+3b)}$ and $C_2^{(4b)}$, as we
expected, and the well-known result \cite{Veltman} is recovered
\begin{equation}
\Delta \rho \simeq \frac{3g^2}{64\pi^2}\frac{m_t^2}{M_W^2}.
\end{equation}
Thus we have confirmed the validity of our method. It is interesting to note
that the contributions of the $t_{<}$ integration (single line of t) to the
effective action yield contributions to the $\Delta \rho$, denoted by
$C_2^{(3c)}$ and $C_2^{(4a)}$, which are relatively suppressed by powers of
$r$.  It should also be noticed that the main contribution to the $\Delta
\rho$ turns out to come from the $C_2^{(3a+3b)}$ alone,  the result of the
integration of higher momentum parts of t and b quarks. Namely, the
Wilsonian renormalization alone gives almost correct answer of the
observable $\Delta \rho$.

(b) Doublet ($\tilde{u}$,$\tilde{d}$) in MSSM

We briefly discuss the contribution of the pair of superpartners,
($\tilde{u}$, $\tilde{d}$), of light quarks u and d ($m_u,m_d \ll M_W$) as a
typical example of the case where the decoupling of new physics contribution
is seen.
In this case the roles of $h$ and $l$  particles are played by $h$:
$\tilde{u}$,$\tilde{d}$, $l$: SU(2) gauge bosons $W^{+}, W^{-}, W^3$,
respectively. Thus,
following our method to obtain $L_{eff}$, both higher and lower momentum
parts
of ($\tilde{u}$, $\tilde{d}$) should be integrated out and the sum of
integrations just provides the final result of $\Delta  \rho$. So our
purpose here is just to see the relative importance of $C_{2(<)}$ and
$C_{2(>)}$, the contributions of single and double lines of ($\tilde{u}$,
$\tilde{d}$), respectively.
Ignoring left-right mixing of squarks, for simplicity, we get
\begin{eqnarray}
C_{2(>)} &=& \frac{3g^2}{64\pi^2}\frac{1}{M_W^2} \{m_{\tilde{u}}^2 +
m_{\tilde{d}}^2
-\frac{2m_{\tilde{u}}^2m_{\tilde{d}}^2}{m_{\tilde{u}}^2 -
m_{\tilde{d}}^2}ln\frac{m_{\tilde{u}}^2}{m_{\tilde{d}}^2} -
\frac{(m_{\tilde{u}}^2 -
m_{\tilde{d}}^2)^2}{3m_{\tilde{u}}^{4}m_{\tilde{d}}^{4}} \mu^{6} \},
\nonumber \\
C_{2(<)} &=& \frac{g^2}{64\pi^2}\frac{1}{M_W^2} \frac{(m_{\tilde{u}}^2 -
m_{\tilde{d}}^2)^2}{m_{\tilde{u}}^{4}m_{\tilde{d}}^{4}} \mu^{6},
\end{eqnarray}
and
\begin{eqnarray}
\Delta \rho &=& \frac{1}{M_W^2}(C_{2(>)} + C_{2(<)}) \nonumber \\
            &=& \frac{3g^2}{64\pi^2}\frac{1}{M_W^2} \{m_{\tilde{u}}^2 +
m_{\tilde{d}}^2
-\frac{2m_{\tilde{u}}^2m_{\tilde{d}}^2}{m_{\tilde{u}}^2 -
m_{\tilde{d}}^2}ln\frac{m_{\tilde{u}}^2}{m_{\tilde{d}}^2} \} ,
\end{eqnarray}
thus recovering the known one \cite{Lim}. We note that $m_{\tilde{u}}^2 =
M_{SUSY}^2 + m_u^2$ and $m_{\tilde{d}}^2 = M_{SUSY}^2 + m_d^2$ ($M_{SUSY}$:
SUSY breaking scale), and under $m_u, m_d \ll M_{SUSY}$ the $\Delta \rho$ is
suppressed by $1/M_{SUSY}^2$ as $\Delta \rho \simeq
\frac{g^2}{64\pi^2}\frac{(m_u^2 - m_d^2)^2}
{M_W^{2} M_{SUSY}^2}$: decoupling occurs. Or we may just say $S_{eff} \simeq
0$
 (except for some renormalization effects to the original action), as
expected. We learn  that the main contribution comes from $C_{2(>)}$,  again
from the  Wilsonian renormalization alone.

\vspace*{0.5 cm}
\leftline{\bf 4. Full effective action}

So far we have retained only the operators in $S_{eff}$ which are relevant
for a specific observable, i.e. $\Delta \rho$. We finally generalize the
argument
and present a systematic way to construct full effective action $S_{eff}$,
obtained by the "integration" of heavy particles. In this article we  take
the Standard Model with one quark generation (t, b) as the typical case, and
will construct full low energy effective action $S_{eff}$ which stems from
the "integration" of $t_{>}$, $t_{<}$ and $b_{>}$ fields. The effective
action $S_{eff}$  turns out to be given as a finite sum of operators
with respect to SU(2) gauge bosons $W^{+}_{\mu}, W^{-}_{\mu}, W^{3}_{\mu}$,
U(1) gauge boson
$B_{\mu}$, Higgs doublet $\phi$ and $b_{<}$, provided only non-decoupling
effects are kept. The non-decoupling effects will be extracted by taking a
formal
limit of $m_t, \ \mu \rightarrow \infty$, keeping the ratio $r = \mu^{2}
/m_t^{2}$ a small constant ($r \ll 1$).

  Here we take the way to directly perform path-integral of  fermionic
fields,
as  the original lagrangian is quadratic in the fermionic fields,
\begin{eqnarray}
L &=& \pmatrix{\bar{t} & \bar{b}}\{ i \partial^{\mu}\gamma_{\mu} -
\pmatrix{m_t & 0 \cr
         0   & m_b} \nonumber \\
  &+& [(gW_{\mu}^i \frac{\sigma^i}{2} + \frac{1}{6}g' B_{\mu})\gamma^{\mu} L
+ g' \pmatrix{\frac{2}{3} & 0 \cr
                0         & -\frac{1}{3}}
 B_{\mu} \gamma^{\mu} R] \}
\pmatrix{t\cr
         b},
\end{eqnarray}
where we have ignored the Higgs doublet $\phi$ except for the role of giving
quarks their masses (it will be finally recovered in the discussion below),
and $L$ and $R$ are chiral-projection operators.
We perform path-integrals of $t_{>}$, $b_{>}$ and $t_{<}$ fields, though we
will eventually see that the contribution of $t_{<}$ integration is less
important.

The effective action $S_{eff}$ is given as
\begin{equation}
S_{eff} = S_{eff(>)} + S_{eff(<)},
\end{equation}
where $S_{eff(>)}$ and $S_{eff(<)}$ denote the  contributions due to the
path-integrals of $t_{>}$, $b_{>}$ and $t_{<}$, respectively. First,
$S_{eff(>)}$ is given as the  $Tr \ ln$ of an operator in the quadratic form
of $t_{>}$, $b_{>}$ fields in the original lagrangian, which reads, after
ignoring a constant term,

\begin{eqnarray}
S_{eff(>)} \simeq i Tr_{(>)} \sum_{n=2}^{4} \frac{1}{n}
\{ \pmatrix{\frac{i}{i\partial^{\mu}\gamma_{\mu} - m_t} & 0\cr
         0   & \frac{i}{i\partial^{\mu}\gamma_{\mu} - m_b}}
i[(g W_{\mu}^i \frac{\sigma^i}{2} + \frac{1}{6}g' B_{\mu})\gamma^{\mu} L
\nonumber \\
+ g' \pmatrix{\frac{2}{3} & 0\cr
                0         & -\frac{1}{3}}
 B_{\mu} \gamma^{\mu} R]\}^n,
\end{eqnarray}
where $Tr_{(>)}$ means that in the fermion propagators, $ S_F(x_1,x_2) =
\int \frac{d^{4}k}{(2\pi)^4}\frac{i e^{-ik\cdot
(x_{1}-x_{2})}}{k^{\mu}\gamma_{\mu} - m_t}$ etc., only higher momentum part
$|k| \ge \mu$ should be integrated. Strictly speaking, this condition may
not be satisfied by all propagators
in the internal lines, if the momenta of external gauge bosons are taken
into account. But as long as
the external momenta are low, $|p^2| \le \mu^2$, the violation of the
condition will
occur
only at a small shell of internal momentum $k^{\mu}$, and we can just
perform the integral of $k^{\mu}$ at the range  $|k| \ge \mu$.
It should be noticed that the summation of $n$ ends up at $n = 4$ (gauge
boson 4-point function). This is because the operators with $n > 4$ is
suppressed by $1/m_t$ and/or $1/\mu$ and disappears at the formal limit,
i.e. decoupled from the low energy theory. Further (Taylor-) expanding the
gauge fields around a space-time point, we will get
operators with derivatives. Let the number of the derivative be $m$.
Then actually only operators satisfying $n + m \leq 4$ survive under the
formal limit as the
candidate of possible non-decoupling contributions. So we end up with a
finite number of local operators, as we expected. Among the remaining ones,
some operators are accompanied by divergent coefficient functions and just
contribute to the renormalizations of existing operators in the original
lagrangian. What we are really interested in are the operators which do  not
exist in the original lagrangian whose coefficient functions should have all
informations of the genuine non-decoupling contributions and are all finite.
The problem of breakdown of local gauge invariance due to U.V. cutoff
$\Lambda$, therefore, does not appear in these coefficients. In the case of
$n = 2$ (gauge boson 2-point functions), it is known
that there exist three independent operators with finite coefficients which
are  related to the S, T and U parameters \cite{Peskin}. In the case of
gauge boson 3-point functions ($n = 3$) it turns out that there appears
additional four
independent operators to describe the non-decoupling effects
\cite{Inami},\cite{Appelquist}. As for the case of $n = 4$, we have not
worked out how many additional operators exist. For instance, $\Delta \rho$
is essentially same as the T parameter, and is understood as one example of
such finite coefficient functions of $S_{eff}$. Next, the path-integral of
$t_{<}$ can be done by completing a square in the relevant part of the
original lagrangian. The result reads as
\begin{eqnarray}
S_{eff(<)} &\simeq& i Tr_{(<)} \sum_{n=2}^{4} \frac{1}{n}
\{ \frac{i}{i\partial^{\mu}\gamma_{\mu} - m_t}
i[(\frac{g}{2}W_{\mu}^3 + \frac{g'}{6} B_{\mu})\gamma^{\mu} L
+ \frac{2g'}{3} B_{\mu} \gamma^{\mu} R] \}^n \nonumber \\
         &+& \int d^{4}x \frac{g^2}{2} \bar{b}W_{\mu}^{-}\gamma^{\mu}
\frac{i\gamma^{\lambda}\partial_{\lambda}}{\Box + m_t^2}
W_{\nu}^{+}\gamma^{\nu}L b ,     \end{eqnarray}
where $Tr_{(<)}$ means that in the t quark propagator only lower momentum
part $|k| \leq \mu$ should be integrated. These contributions are less
important compared with $S_{eff(>)}$, as they are
relatively suppressed by the powers of $r = \mu^2/m_t^2$, coming from the
suppression of t quark propagator by $1/m_t$ and the limitation of phase
space, $|k| \leq \mu$. This situation can be explicitly confirmed in the
case of
$\Delta \rho$, by comparing $C_2^{(3a+3b)}$ with either $C_2^{(3c)}$ or
$C_2^{(4a)}$, corresponding to the contribution shown in the first or the
second lines of eq.(16). Thus combining $S_{eff(>)}$ with the remaining part
of the original action with respect to light fields including b quark
(with suitable renormalizations), we get whole theory in low-energies.

The operators which do  not exist in the original lagrangian, the genuine
effects of heavy particle, can be clearly extracted, once we introduce Higgs
doublet
field. All quantum corrections should be written by local gauge invariant
operators, if the VEV is replaced by the original Higgs doublet, and
 the operators which do  not exist in the original lagrangian should be
written as operators with higher mass dimension ($d > 4$), i.e. by
irrelevant operators. So these operators have automatically finite
coefficient functions and can be clearly separated  from
marginal or relevant operators with $d \leq 4$. One problem of this project
is that infinite number of higher dimensional operators may potentially
 participate in describing the non-decoupling effects of t quark.
 This is because once we have a gauge invariant operator we may trivially
extend it by putting powers of $\phi^{\dagger}\phi$ ($\phi$: Higgs doublet).
These
 extended operators may be accompanied by additional inverse powers of $v$,
the VEV of the Higgs field. After replacing $\phi$ by its VEV, however,
$\phi^{\dagger}\phi/v^2$ just gives 1 and all of these operators play the
role to describe the non-decoupling effects on an equal footing. (In new
physics theories where decoupling of heavy particles are expected, such
higher dimensional operators are suppressed by the inverse powers of a new
gauge invariant large mass scale, say $M$, as $\phi^{\dagger}\phi/M^2$ and
the problem does not exist). This problem may be evaded  if we utilize
non-linear realization of the Higgs doublet, $\phi = U \pmatrix{0 \cr v}$
with
 $U = exp(iG^{a}\sigma^{a}/2v)$. For simplicity we have ignored the physical
Higgs field $h^0$, but it will be recovered shortly. Now powers of
$(\phi^{\dagger}\phi) /v^2$ just gives 1 and the problem disappears. Thus
our procedure is just
 to replace the quark mass matrix as
\begin{equation}
\pmatrix{m_t & 0 \cr
         0   & m_b}
\rightarrow
U \pmatrix{m_t & 0 \cr
         0   & m_b} R
+
\pmatrix{m_t & 0 \cr
         0   & m_b} U^{\dagger} L.
\end{equation}
After a change of variables of path-integration utilizing a field dependent
local SU(2)$_L$ transformation,
\begin{equation}
\pmatrix{t \cr
         b   }_{L}
\rightarrow
\pmatrix{t' \cr
           b' }_{L}
 = U^{\dagger}
\pmatrix{t \cr
         b}_{L},
\end{equation}
in order to eliminate $U$ from the replaced quark mass matrix, the
path-integration in terms of the new variables yields the final result of
the effective action,

\begin{eqnarray}
S_{eff} \simeq i Tr_{(>)} \sum_{n=2}^{4} \frac{1}{n}
\{ \pmatrix{\frac{i}{i\partial^{\mu}\gamma_{\mu} - m_t(1+\frac{h^0}{v})} &
0\cr
         0   & \frac{i}{i\partial^{\mu}\gamma_{\mu} - m_b(1+\frac{h^0}{v})}}
i[U^{\dagger}(i D_{\mu} U)\gamma^{\mu} L \nonumber \\
+ \frac{1}{6}g'B_{\mu} \gamma^{\mu}L +  g'\pmatrix{\frac{2}{3} & 0\cr
                0         & -\frac{1}{3}}
 B_{\mu} \gamma^{\mu} R]
\}^n
 - ("relevant" operators),
\end{eqnarray}
where $D_{\mu}$ is SU(2) covariant derivative, $D_{\mu} = \partial_{\mu} -
igW_{\mu}^i \frac{\sigma^i}{2}$, and the contribution from $t_{<}$
integration
has been ignored. In this final result we have recovered the Higgs field
$h^0$
by a simple replacement, $m_q \rightarrow m_q(1+\frac{h^0}{v})$.
Again we may retain only operators, whose mass dimensions do not exceed 4.
As the field $U$ is
dimensionless,
the irrelevant operators (relevant or marginal operators) in the linear
realization of Higgs doublet correspond to operators in which the sum of the
numbers of $U$ field, gauge bosons and derivative  exceeds 4 (does not
exceed 4). The subtraction of $"relevant" operators$ in the above expression
should be understood in such a sense. For instance, focusing on the
operators
in $S_{eff}$ which is quadratic in SU(2) gauge fields ($n = 2$) and
without any derivative, we get two types of
operators.  One is $Tr [(D_{\mu}U)^{\dagger}D^{\mu}U]$ ($Tr$ should be
understood to stand for the trace of matrix elements ). This is a "relevant"
operator and
should be subtracted. Another one is $Tr [(U^{\dagger}i D^{\mu} U)\sigma_{3}
(U^{\dagger}i D_{\mu} U)\sigma_{3}]$, which is clearly an "irrelevant "
operator. We can easily check that setting $U = 1$ the operator gives a
gauge boson mass-squared term corresponding to $\Delta \rho$,
$W^{1\mu}W^{1}_{\mu} + W^{2\mu}W^{2}_{\mu} - W^{3\mu}W^{3}_{\mu}$. The
coefficient  of this operator is proportional to  \begin{equation}
i \int \frac{d^{4}k}{(2\pi)^4} \frac{(m_{t}^{2}-m_{b}^{2})^{2}k^2 }
{(k^2 -m_{t}^{2})^2(k^2 -m_{b}^{2})^2},
\end{equation}
which exactly gives the formula of $\Delta \rho$. The presence of the factor
$(m_{t}^{2}-m_{b}^{2})^{2}$ is what we expect from a symmetry argument
\cite{Yamada}. $\Delta \rho$ or T-parameter is an observable which behaves
as a 5-plet repr. of SU(2) and needs the $m_{t}^{2}-m_{b}^{2}$ factor,
behaving as a triplet
of SU(2), twice.

We finally ask a question, to what extent the result of Wilsonian
renormalization alone is close to the exact result of $S_{eff}$. We have
seen in this
article that if we take a specific observable $\Delta \rho$, the
$S_{eff(>)}$ alone gives almost correct
answers in two examples of gauge models. The dominance of $S_{eff(>)}$,
however, is not always true. In the case of $\Delta \rho$, the dominance of
higher momentum
part (double-line intermediate states) can be understood as a consequence of
the fact that  $\Delta \rho$ is concerned with gauge boson mass-squared,
which
of
course has mass dimension 2. Therefore if we consider dimensionless
observables, we expect to have the same order of logarithmic contributions
both from $S_{eff(>)}$ and from the original action with respect to low
momentum part of light particles. For instance the S-parameter \cite{Peskin}
parametrizes a mixing between the field strengths of SU(2) and U(1) gauge
bosons, and is
dimensionless observable.
In the case of (t, b) contribution, a part of S-parameter (proportional to
B-L charge) behaves as
$ln (m_t/m_b)$. In our approach,  $S_{eff(>)}$ and the low momentum part of
b quark give contributions to the S-parameter behaving as $ln (m_t/\mu)$ and
$ln (\mu/m_b)$,
respectively, which are comparable to each other.

\vspace*{0.5 cm}
\leftline{\bf 5. Concluding remarks}

In conclusion, we have proposed a systematic method to
obtain effective low energy action for light particles in a system
where both heavy and light particles coexist with a close relation, like the
case of  (t, b) system. Thanks to Wilsonian renormalization, we could get
effective action with finite number of local operators describing possible
non-decoupling effects of heavy particles. The validity of the method was
first checked by the computation of the $\rho$ - parameter by use of the
derived relevant part of the effective action. Then we constructed
full effective action with respect to light particles, which contain all
informations of the non-decoupling effects,  by directly
performing path-integrals of fermionic fields. A useful way to extract
genuine contributions of heavy
particles, described by irrelevant operators, was discussed to exist
once
we use non-linear realization of Higgs doublet, though we were not
interested in the Green functions of Higgs field themselves. As the
irrelevant
operators have finite coefficient functions, we did not suffer from the
explicit breakdown of local
gauge invariance due to the regularization by momentum cutoff.
We have seen that only the path-integration of higher momentum parts of
fields, i.e. Wilsonian renormalization alone, always gives practically
correct answer for $S_{eff}$. In this work we did direct path-integration
only for fermionic fields. If we also perform path-integrations of the
higher momentum parts of gauge
bosons and Higgs field, we will get an effective action including operators
of fermionic fields only, which are out of scope of the present work. We
hope
that the procedure proposed in this paper is useful in deriving full
low-energy effective action with finite
number of local operators (healthy action !) starting from an
original lagrangian,  in any system with both heavy and light particles.

\noindent {\bf Acknowledgment}

 The authors would like to thank S. Hasegawa for his collaboration in early
stage of this work. Thanks are also due to T. Inami and H. Sonoda for
helpful discussions.

 This work has been supported by Grant-in-Aid for Scientific Research
(09640361)
from the Ministry of Education, Science and Calture, Japan.

\end{document}